\documentclass[conference]{IEEEtran}
\IEEEoverridecommandlockouts

\usepackage{cite}
\usepackage{amsmath,amssymb,amsfonts}
\usepackage{algorithmic}
\usepackage{graphicx}
\usepackage{textcomp}
\usepackage{xcolor}
\def\BibTeX{{\rm B\kern-.05em{\sc i\kern-.025em b}\kern-.08em
    T\kern-.1667em\lower.7ex\hbox{E}\kern-.125emX}}

\usepackage{tikz}

\usepackage{enumitem}
\usepackage{textcomp}
\usepackage{listings} % For code listings
\usepackage{algorithm}
\usepackage{makecell}
\usepackage{multirow}
\usepackage{booktabs}
\usepackage{pifont}
\usepackage{array}
\newcommand{\cmark}{\ding{51}}
\newcommand{\xmark}{\ding{55}}
\usepackage{arydshln}
\usepackage{booktabs, tabularx}

\usepackage{filecontents}

\definecolor{codegreen}{rgb}{0,0.6,0}
\definecolor{codegray}{rgb}{0.5,0.5,0.5}
\definecolor{codepurple}{rgb}{0.58,0,0.82}
\definecolor{backcolour}{rgb}{0.95,0.95,0.92}
\definecolor{mygreen}{RGB}{0,128,0}
\definecolor{codeblue}{rgb}{0,0.6,0.8}

\makeatletter
\newcommand{\linebreakand}{%
 \end{@IEEEauthorhalign}
 \hfill\mbox{}\par
 \mbox{}\hfill\begin{@IEEEauthorhalign}
}
\makeatother

\begin{document}

\title{EMO: Energy Efficiency Modeling and Optimization for AI Workloads}

\author{

\IEEEauthorblockN{Jiyu Luo}
\IEEEauthorblockA{\textit{University of Science and Technology} \\ \textit{of China} \\
Hefei, China \\
luojiyu@mail.ustc.edu.cn}
\and
\IEEEauthorblockN{Shaoyu Chen}
\IEEEauthorblockA{\textit{University of Science and Technology} \\ \textit{of China} \\
Hefei, China \\
rktrem@mail.ustc.edu.cn}
\and
\IEEEauthorblockN{Jingwei Sun}
\IEEEauthorblockA{\textit{University of Science and Technology} \\ \textit{of China} \\
Hefei, China \\
sunjw@ustc.edu.cn}
\linebreakand
\IEEEauthorblockN{Shengcai Liu}
\IEEEauthorblockA{\textit{Southern University of Science and}\\ \textit{Technology} \\
Shenzhen, China \\
\textit{Guangdong Provincial Key Laboratory of }\\ \textit{Brain-Inspired Intelligent Computation} \\
Shenzhen, China \\
liusc3@sustech.edu.cn}
\and
\IEEEauthorblockN{Ke Tang}
\IEEEauthorblockA{\textit{Southern University of Science and}\\ \textit{Technology} \\
Shenzhen, China \\
\textit{Guangdong Provincial Key Laboratory of }\\ \textit{Brain-Inspired Intelligent Computation} \\
Shenzhen, China \\
tangk3@sustech.edu.cn}
\and
\IEEEauthorblockN{Guangzhong Sun}
\IEEEauthorblockA{\textit{University of Science and Technology} \\ \textit{of China} \\
Hefei, China \\
gzsun@ustc.edu.cn}
}

\maketitle

\begin{abstract}

The massive energy consumption of GPU-accelerated AI workloads challenges sustainable computing. We observe that execution asynchrony (e.g., CPU-GPU, concurrent streams, multi-GPU) creates slack, allowing non-critical kernels to run at lower frequencies to save energy without impacting end-to-end latency. However, existing approaches fail to simultaneously achieve workload generality and fine-grained slack discovery, while high-fidelity modeling incurs prohibitive overhead.

We present EMO, a lightweight framework exploiting these fine-grained opportunities. First, to identify \textit{where} to optimize, EMO constructs a low-level dependency graph capturing asynchrony and performs what-if timing analysis to precisely identify slack windows. Second, to determine \textit{how} to optimize, EMO introduces dependency-aware kernel packing. It aggregates kernels to preserve critical paths while collapsing redundant details, enabling high-fidelity latency-energy modeling with minimal profiling cost. Finally, EMO combines graph analysis and pack-level models to formulate energy optimization as a constrained combinatorial problem, efficiently solving for optimal frequency policies under given latency targets.
Evaluations show EMO reduces energy consumption by 15\%--28\% with only 2\%--5\% performance loss and negligible overhead.

\end{abstract}


\section{Introduction}

The rapid advancement of deep learning has dramatically increased computational power demands across increasingly diverse AI workloads, including training and inference~\cite{yang2024qwen2technicalreport,deepseekai2024deepseekv2strongeconomicalefficient,NEURIPS2025_2e99eb4e,wu2026trainmovingedgeonlineverified,pmlr-v267-lu25g}, leading to widespread adoption of GPUs and specialized accelerators in both cloud and on-premise datacenters.
As models and datasets continue to scale, the resulting surge in energy consumption has become a critical concern. 
Studies indicate that training large language models such as GPT-3 can consume energy on the order of gigawatt-hours (GWh)
~\cite{10363447,jegham2025hungryaibenchmarkingenergy,patterson2021carbonemissionslargeneural,rodriguez_evaluating_2024}, driving energy costs to represent a substantial portion of total AI infrastructure expenditure. This escalating power consumption not only inflates operational costs but also constrains scalability and sustainability objectives~\cite{cbreGlobalData,tschand_mlperf_2025}.

Dynamic Voltage and Frequency Scaling (DVFS) is widely supported on modern accelerators, providing trade-offs between performance and energy. Hardware-integrated techniques such as CRISP~\cite{7856605}, PCSTALL~\cite{bharadwaj_predict_2023}, and ReGate~\cite{regate} enable nanosecond-level responsiveness but rely on customized hardware support or simulators, limiting their applicability. In contrast, software runtimes expose DVFS through vendor APIs and offer superior flexibility and portability, making them the predominant approach for energy optimization in production systems~\cite{yue_everest_2025,zhang_improving_2024}.

Existing software approaches generally fall into two categories: model-free and model-based. Model-free methods~\cite{you_zeus_nodate,qiu_power-aware_nodate,wang_model-free_2024,zhang_improving_2024,yue_everest_2025} avoid detailed modeling by using online metrics (e.g., utilization) to reactively adjust frequencies. Model-based methods~\cite{choi_envpipe_nodate,chung_reducing_2024,kakolyris_throttllem_2025,wang_using_2025,geng_powerlens_2024} construct detailed performance-energy models to determine optimal frequency scaling policies. 
However, a fundamental limitation persists across both categories. Neither approach effectively captures fine-grained slack of the AI execution graph, leading to missed optimization opportunities. 
 We identify two fundamental challenges in achieving optimal energy efficiency:

\begin{figure}[!t]
\centering
\includegraphics[width = 0.49\textwidth]{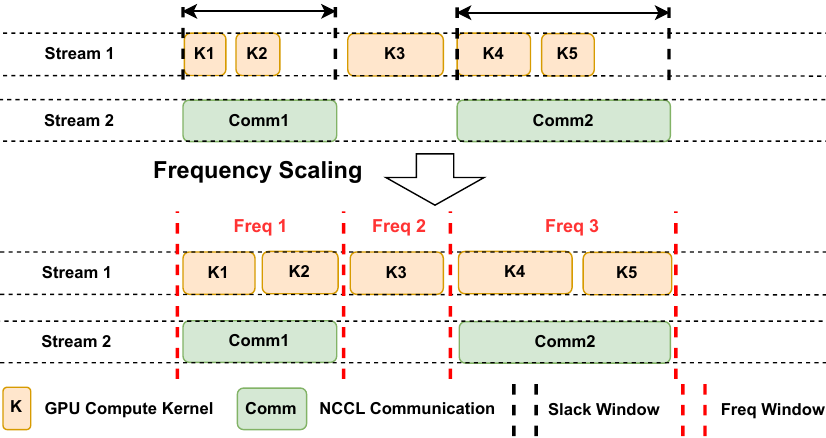}
\caption{Asynchronous execution creates slack where non-critical kernels can run at lower frequencies to save energy without end-to-end performance loss.}
\label{async}
\end{figure}

\textbf{Challenge 1: The Dilemma of Slack Discovery: Generality vs. Granularity.}
Modern AI workloads exhibit complex, asynchronous parallelism where tasks span CPUs, GPUs, and network interconnects concurrently. This asynchrony creates \textit{execution slack}, defined as time windows where kernels are not on the critical path of the workload. As illustrated in Figure~\ref{async}, we can decrease the frequency of a kernel within this slack to save energy. This does not affect the end-to-end iteration time because the delay is hidden by other concurrent tasks.

However, discovering this slack presents a fundamental dilemma between generality and granularity. Existing model-free approaches are workload-agnostic and highly general, but they rely on coarse-grained metrics like average GPU utilization~\cite{yue_everest_2025}. A kernel might exhibit high utilization but still be non-critical because it is waiting for a dependency (e.g., network communication). Consequently, they either miss savings or aggressively scale down critical kernels, causing performance degradation. Conversely, existing model-based approaches~\cite{chung_reducing_2024} can identify fine-grained slack, but they strongly rely on specific workload semantics (e.g. microbatches, pipeline stages). This heavy reliance inherently sacrifices workload generality, rendering them ineffective for arbitrary or emerging AI models.

\textbf{Challenge 2: The Tension between Modeling Fidelity and Profiling Overhead.}
Exploiting execution slack requires fine-grained DVFS decisions at per-kernel granularity or millisecond-scale to distinguish critical operations. It relies on a high-fidelity latency-energy model that predicts how changing frequency affects latency and energy for specific kernels under specific dependency contexts.

However, obtaining this fidelity creates a fundamental conflict with profiling overhead. Accurate modeling typically demands exhaustive runtime profiling~\cite{guerreiro_gpgpu_2018,wang_gpgpu_2020} or pre-built lookup tables~\cite{stojkovic_dynamollm_2025} to map variations across all frequency settings. Such detailed characterization incurs prohibitive costs, as any change in workload, input size, or hardware configuration necessitates repeating the process. This creates a tension: while high-fidelity modeling is essential for identifying slack, the associated profiling overhead renders it impractical for dynamic production environments.

To address these challenges, this paper presents \textbf{EMO}, a lightweight \textbf{E}nergy \textbf{M}odeling and \textbf{O}ptimizing framework that systematically discovers and exploits fine-grained energy optimization opportunities in AI workloads. Our design is built on two design insights: (1) optimization potential lies within low-level execution dependencies, and (2) effective latency-energy modeling requires a proper granularity that balances fidelity with runtime efficiency.

EMO operates in three phases. 
First, it abstracts the workload execution into a fine-grained \textbf{Execution Graph}. This graph captures GPU compute events, CPU events, and communication primitives, along with their happens-before constraints. 
Second, to reduce modeling overhead, EMO employs \textbf{dependency-aware kernel packing}. Instead of profiling every GPU kernel, EMO aggregates kernels into representative packs that preserve critical dependencies while collapsing redundant scheduling details. EMO combines lightweight profiling with microbenchmark-based predictors to build accurate latency-energy models for these packs. 
Finally, EMO integrates these pack-level models into the execution graph to perform a holistic what-if analysis. 
It formulates the energy minimization task as a constrained combinatorial optimization problem and employs a dynamic programming algorithm to efficiently derive the optimal frequency policy within a performance budget.

Our contributions are as follows:
\begin{itemize}[topsep=3pt, left=5pt, itemsep=0pt,parsep=0pt]
  \item We propose a general and lightweight framework for energy modeling and optimization. It performs what-if analysis on low-level dependency graphs. This approach captures comprehensive asynchronous behaviors, enabling the precise identification of execution slack and optimization opportunities that prior methods miss.
  \item We introduce dependency-aware kernel packing techniques that substantially reduce profiling overhead for fine-grained modeling. By combining profiling results with microbench-based performance prediction, we enable accurate and efficient estimation of the latency and energy consumption for each pack under various frequency settings.
  \item We formulate energy efficiency optimization as a combinatorial optimization problem and design an efficient algorithm that yields near-optimal solutions. Evaluations show EMO reduces energy consumption by 15\%--28\% with only 2\%--5\% performance loss and negligible overhead.
\end{itemize}

\begin{table*}[htbp]
\centering
\caption{Comparison of various energy optimization methods.}
\renewcommand{\arraystretch}{1.2}
\begin{tabular}{
    l
    >{\centering\arraybackslash}p{2.4cm}
    >{\centering\arraybackslash}p{2.4cm}
    >{\centering\arraybackslash}p{2.1cm}
    >{\centering\arraybackslash}p{3cm}
    >{\centering\arraybackslash}p{2.4cm}
}
\toprule
\textbf{Related Work} &
\makecell{\textbf{Slack Discovery}\\\textbf{Granularity}} &
\makecell{\textbf{Workload}\\\textbf{Generality}} &
\makecell{\textbf{Profiling}\\\textbf{Overhead}} &
\makecell{\textbf{Online Adaptation}\\\textbf{Overhead}} &
\textbf{Platform} \\
\midrule
MF-GPOEO~\cite{wang_model-free_2024}      & Coarse & \cmark & Low      & High     & NVIDIA           \\
ZEUS~\cite{you_zeus_nodate}          & Coarse & \cmark & Low    & High      & NVIDIA           \\
EnvPipe~\cite{choi_envpipe_nodate}       & Fine   & \xmark & High  & Low      & NVIDIA           \\
PERSEUS~\cite{chung_reducing_2024}       & Fine   & \xmark & High  & Low      & NVIDIA           \\
GEEPAFS~\cite{zhang_improving_2024}       & Coarse & \cmark & Low   & High     & NVIDIA           \\
$\mu$-Serve~\cite{qiu_power-aware_nodate}       & Coarse & \cmark & Low   & High     & NVIDIA           \\
EVeREST~\cite{yue_everest_2025}       & Coarse & \cmark & Low & High  & NVIDIA, AMD      \\
throttLL’eM~\cite{kakolyris_throttllem_2025}   & Fine   & \xmark & High     & Low      & NVIDIA           \\
DynamoLLM~\cite{stojkovic_dynamollm_2025}  & Fine  & \xmark & High     & Low      & NVIDIA           \\
Wang et. al.~\cite{wang_using_2025}          & Coarse & \cmark & Moderate & Low      & Ascend       \\
EMO (Ours)              & Fine   & \cmark  & Low & Low      & NVIDIA           \\
\bottomrule
\end{tabular}
\label{tab:energy_optimization_methods}
\end{table*}

\section{Related Work}

\textbf{Modeling Energy Efficiency.}
Accurate optimization of energy efficiency on GPUs requires detailed knowledge of workload and performance characteristics under different frequency configurations. 
Guerreiro et al.~\cite{guerreiro_gpgpu_2018,guerreiro_modeling_2019,wang_gpgpu_2020,alavani_program_2023,braun_simple_2021,wang_gpgpu_2020-1} construct analytical or machine learning models by collecting runtime performance metrics to predict kernel performance and energy consumption.
For large-scale AI workloads, the profiling overhead associated with collecting fine-grained data becomes prohibitive. Some approaches sidestep runtime profiling by leveraging static information~\cite{sk_powertrain_2024,wang_dso_2024} to predict performance characteristics without execution. Lookup table approaches~\cite{stojkovic_dynamollm_2025,chung_reducing_2024} record kernel or forward/backward pass latency and energy consumption offline under different frequencies to reduce profiling overhead. However, hardware limitations complicate fine-grained energy measurement at millisecond granularity, often requiring hundreds of kernel executions to obtain accurate energy readings~\cite{guerreiro_gpgpu_2018}. 
Moreover, these profiling efforts must be regenerated when workloads change, incurring substantial additional overhead.

\textbf{Optimizing Energy Consumption.}
We categorize energy optimization methods into two categories: model-free and model-based approaches. Model-free methods typically employ feedback-based frequency tuning~\cite{wang_model-free_2024,zhang_improving_2024,yue_everest_2025,you_zeus_nodate,qiu_power-aware_nodate} and operate on coarse-grained system metrics such as utilization. 
Although recent works like $\mu$-Serve~\cite{qiu_power-aware_nodate} introduce operator-level frequency sensitivity modeling, they essentially remain feedback-driven approaches that cannot capture fine-grained execution slack. Consequently, the lack of comprehensive dependency modeling prevents them from accurately identifying optimization opportunities, resulting in suboptimal energy efficiency. Additionally, feedback-based tuning often results in slow or incomplete convergence, requiring extensive online exploration that disrupts normal runtime execution.

Model-based approaches~\cite{stojkovic_dynamollm_2025,choi_envpipe_nodate,chung_reducing_2024,kakolyris_throttllem_2025,wang_using_2025,ali_performance-aware_2023,wang_dynamic_2022} leverage pre-constructed models to formulate energy optimization as a structured search problem.
Wang et al.~\cite{wang_using_2025} collect power consumption data to build analytical models that predict kernel energy and execution time, then employ genetic algorithms to search for optimal frequency policies.
Beyond kernel-level modeling, some studies focus on workload-specific analysis to uncover optimization opportunities. Works such as Perseus~\cite{choi_envpipe_nodate,chung_reducing_2024} exploit asynchrony in pipeline-parallel execution to mask frequency scaling latencies, proposing two types of ``energy bloat'' that correspond to inter-GPU asynchrony in our taxonomy.
Similarly, works such as throttLL’eM~\cite{stojkovic_dynamollm_2025,kakolyris_throttllem_2025,stojkovic_tapas_2025} perform in-depth analysis of LLM serving workloads to optimize energy consumption. Although such workload-specific analysis enables deeper optimization, it inherently limits generalizability. Currently, no existing approach provides fine-grained optimization capabilities that generalize across arbitrary workloads.

Table~\ref{tab:energy_optimization_methods} provides a qualitative comparison of representative energy optimization techniques across several dimensions: (1) the granularity of their slack discovery, (2) their workload generality, (3) profiling overhead, (4) online adaptation overhead, and (5) supported platforms. Our approach, EMO, establishes a new balance: achieving fine-grained, workload-agnostic energy optimization with minimal profiling and online adaptation overhead, making it practical for deployment in diverse AI computing environments.

\section{Key Insights}
\label{sec:key-insights}
Our design is guided by two key insights that address the fundamental challenges of optimizing end-to-end energy efficiency for diverse AI workloads.

\begin{figure}[!htbp]
\centering
\includegraphics[width = 0.49\textwidth]{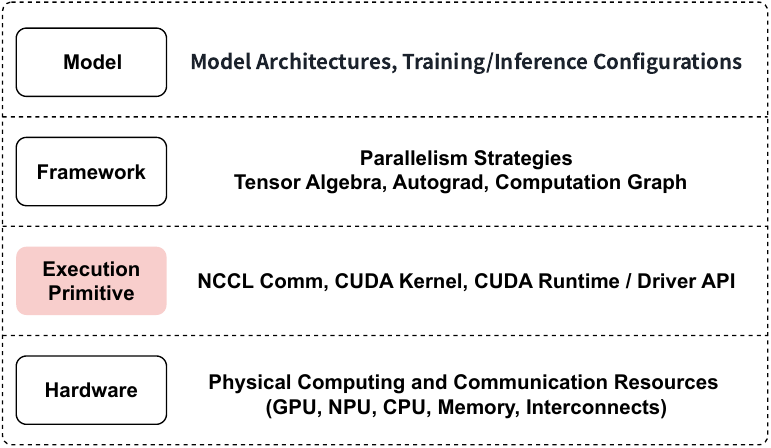}
\caption{Abstraction Layers in Modern AI Systems}
\label{lowlevel}
\end{figure}

\subsection{Workload-Agnostic Slack Discovery via Execution Graph}

To identify where energy can be saved without hurting performance, we must first understand the complex execution flow of AI workloads. Modern AI systems are not monolithic; they are orchestrated across heterogeneous resources involving GPU computation, interconnect communication, and diverse CPU operations ranging from control logic to compute-heavy offloading. This interaction creates inherent asynchrony, which we categorize into three types:

\begin{enumerate}[label=(\arabic*), topsep=1pt, left=1pt, itemsep=0pt,parsep=0pt]
    \label{async2}
    \item \textbf{CPU-GPU Asynchrony:} CPU and GPU execute tasks independently during memory-constrained offloading or heterogeneous computation~\cite{vellaisamy2025characterizingoptimizingllminference,10.1145/3689031.3696067,273920,10.1145/3689031.3717461}.
    \item \textbf{Intra-GPU Asynchrony:} Communication and computation proceed concurrently within a single GPU, particularly in data parallelism and tensor parallelism implementations~\cite{chang2024flux,wang2024-deepspeed-domino,zhang2025cometfinegrainedcomputationcommunicationoverlapping}.
    \item \textbf{Inter-GPU Asynchrony:} Multiple GPUs perform different tasks or exhibit varying computational intensities, causing load imbalance or straggler effects in pipeline parallelism and heterogeneous environments~\cite{10.1145/3341301.3359646,298557,lin2025understandingstragglerslargemodel}.
\end{enumerate}

Capturing these asynchronous behaviors is crucial for locating execution slack. However, as shown in Figure~\ref{lowlevel}, the choice of abstraction level determines the visibility of these opportunities. High-level abstractions, such as model architectures or framework computation graphs, focus on logical data dependencies but obscure the runtime parallelism (e.g., overlapping communication with computation). Conversely, hardware-level instruction traces provide full detail but are too noisy and expensive to collect for runtime optimization.

We identify the \textbf{Execution Primitive} level as the optimal abstraction layer to break the dilemma between generality and granularity. This level consists of CUDA kernels, runtime API calls, and NCCL operations. Regardless of the upper-level model semantics or frameworks, all workloads ultimately decompose into these fundamental primitives. By constructing an \textbf{Execution Graph} at this level, we can explicitly map the happens-before dependencies between primitives. This graph provides the fine-grained visibility needed to pinpoint execution slack, while remaining workload-agnostic. It enables us to perform precise ``what-if'' analysis to predict how frequency adjustments on specific primitives propagate through the graph and affect end-to-end latency, without relying on any predefined model structures.

\begin{figure}[htbp]
    \centering
    \includegraphics[width=0.49\textwidth]{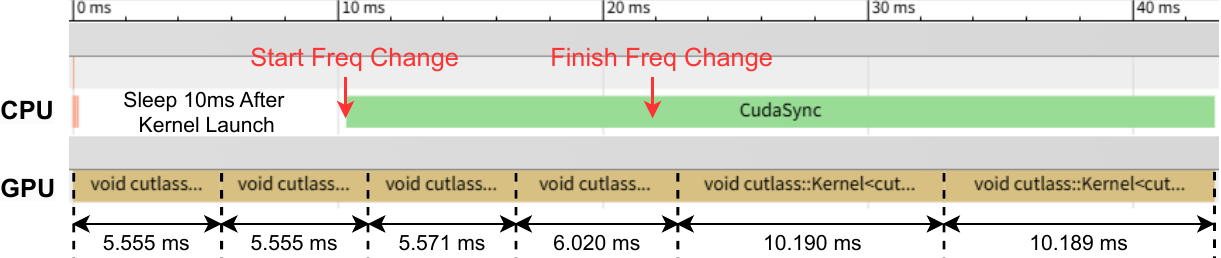} 
    \caption{Execution trace on an RTX 3080 Ti demonstrating the overhead of frequency scaling. Six identical kernels are executed sequentially, and a frequency change from 1800 MHz to 900 MHz is initiated at 10 ms, and taking effect during the fourth kernel.}
    \label{fig:freq_overhead}
\end{figure}

\subsection{Kernel Packing Balances Modeling Fidelity and Overhead}

While the execution graph reveals \textit{where} optimization opportunities exist, the challenge shifts to \textit{how} to efficiently model and select frequencies. As highlighted in Challenge 2, a fundamental tension exists between modeling fidelity and profiling overhead. Characterizing the energy and latency sensitivity of every individual kernel to achieve high fidelity incurs prohibitive runtime costs.

While recent studies~\cite{stojkovic_dynamollm_2025} report that GPU frequency adjustments incur overheads and degrade throughput, our profiling reveals a different hardware reality. As illustrated in Figure~\ref{fig:freq_overhead}, when we execute identical kernels and initiate a frequency change, the adjustment process is entirely asynchronous. It does not block GPU execution or create bubbles, and the actual hardware-level frequency transition occurs with negligible interference to the running kernels.

Although the scaling process is non-blocking, the DVFS command still requires an interval of 5 to 15 milliseconds~\cite{wang_gpgpu_2020} to take effect. 
While we can pre-issue commands to align the actual frequency transition with the start of slack windows, this delay fundamentally limits the control granularity.
Consequently, modeling performance at a sub-millisecond per-kernel resolution is wasteful and provides redundant information that the DVFS controller cannot exploit. 

This observation leads to our second insight: Kernel Packing. To eliminate wasteful profiling, we define a pack as a continuous sequence of GPU compute kernels aggregated to match this control granularity. We specifically isolate compute kernels because they are the primary energy consumers sensitive to frequency scaling, whereas communication~\cite{10818209} and CPU tasks are generally frequency-tolerant. By evaluating at the pack level, EMO reduces the number of entities requiring profiling, collapsing redundant scheduling details.

However, arbitrary packing would obscure the structural insights gained from the execution graph. To resolve this, EMO introduces \textit{dependency-aware} kernel packing in the next section. It aggregates kernels only when such grouping preserves the critical-path relationships identified in the graph.

\begin{figure}[!htbp]
\centering
\includegraphics[width = 0.49\textwidth]{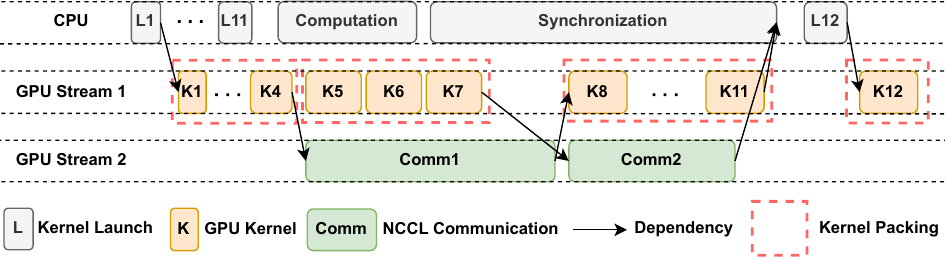}
\caption{A Kernel Packing Example}
\label{packing}
\end{figure}

\section{Design of EMO}

\subsection{Overview}
\begin{figure}[!htbp]
\centering
\includegraphics[width = 0.46\textwidth]{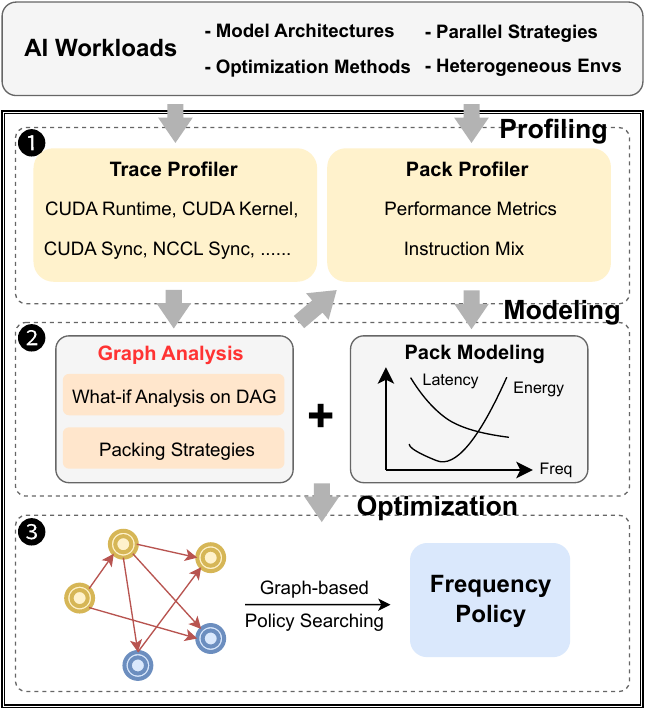}
\caption{Overview of EMO}
\label{workflow}
\end{figure}

Building on the insights discussed above, we introduce EMO, a general and lightweight framework for energy modeling and optimization across diverse AI workloads. As illustrated in Figure~\ref{workflow}, EMO comprises three key components: \ding{202} the \textbf{Profiling} phase, \ding{203} the \textbf{Modeling} phase, and \ding{204} the \textbf{Optimization} phase.

Central to this framework is the \textbf{Graph Analysis} within the modeling phase. For a given AI workload, EMO first employs a trace profiler to capture low-level execution primitives and their dependencies, constructing a Directed Acyclic Graph (DAG) that fully characterizes the workload's asynchronous behavior. 
It then performs a what-if Analysis on the DAG to formally quantify how frequency scaling of GPU compute kernels affects end-to-end latency and total energy consumption. Based on the dependency structure revealed by the DAG, we select appropriate packing strategies to aggregate GPU kernels into coarse-grained packs. 

For a small subset of representative packs, we utilize a lightweight pack profiler to record runtime performance metrics. These metrics are fed into a microbenchmark-based model to predict the latency and energy sensitivity of each pack to frequency changes. Finally, by integrating pack-level modeling with the holistic DAG analysis, EMO asynchronously searches for and applies near-optimal frequency configurations at runtime. This enables continuous online energy efficiency optimization, adapting to workload characteristics without interrupting or significantly impacting execution.

\subsection{Profiling}

Although several mature GPU profilers exist, such as NVIDIA Nsight Systems, Nsight Compute~\cite{ncu,nsys}, PyTorch Profiler~\cite{kineto}, and various simulation-based tools~\cite{daydream,habitat,vtrain}, these solutions do not fully satisfy EMO's requirements for lightweight, flexible-purpose profiling. 
In particular, they lack native support for kernel packing, efficient pack-level metric collection, and low-level execution graph construction. To address these limitations, we implement EMOprofiler directly on top of CUPTI~\cite{cupti}. EMOprofiler consists of two main components: the trace profiler and the pack profiler.

The \textbf{Trace Profiler} captures key execution primitives and their dependencies during program execution. We utilize three categories of primitives to represent all program events as nodes in the graph. First, for \textbf{CPU Events}, we capture all CUDA Runtime APIs via CUPTI's Activity API. We define the interval between two consecutive runtime APIs as a fixed CPU computation event. This design is based on the observation that GPU frequency scaling does not impact CPU latency, and CPU-side logic or computation are inherently captured within these intervals. Second, for \textbf{GPU Kernels}, we record CUDA kernels via CUPTI's Activity API. Finally, regarding \textbf{Communication}, NCCL operations are instrumented via \texttt{LD\_PRELOAD} to record details such as communication domains and participants.

Accurately capturing dependencies as DAG edges is essential for reconstructing asynchronous and overlapping execution, which is critical for understanding the impact of frequency scaling. Similar to the asynchrony categories in the Section \ref{sec:key-insights}, we classify dependencies into three categories:
\begin{enumerate}[label=(\arabic*), topsep=3pt, left=3pt]
    \item \textbf{CPU-GPU}: Explicit dependencies including kernel launch APIs, CUDA stream/event synchronizations, and memory copy APIs.
    \item \textbf{Intra-GPU}: Explicit dependencies where synchronization between streams is established via CUDA EventRecord and enforced with CUDA StreamWaitEvent, enabling control computation and communication overlap.
    \item \textbf{Inter-GPU}: Implicit dependencies where various communication APIs use polling mechanisms to synchronize across devices~\cite{nccl}.
\end{enumerate}

The \textbf{Pack Profiler} enables online collection of pack-level metrics. By intercepting kernel launches via CUPTI Callbacks APIs, EMOprofiler inserts user-range profiling APIs before and after each identified pack, following the packing strategies in Section~\ref{analysis}. Similar to existing work~\cite{wang_gpgpu_2020,guerreiro_gpgpu_2018,guerreiro_modeling_2019}, we collect several key metrics that provide the foundation for subsequent performance and energy modeling, as summarized in Table~\ref{tab:metrics}.

\begin{table}[htbp]
\centering
\caption{GPU Performance Metrics}
\renewcommand{\arraystretch}{1.2}
\begin{tabular}{ll}
\toprule
ID & Description \\
\midrule
Cyc   & SM active cycles \\
Warp  & Avg. active warps per cycle \\
DRR/W   & DRAM bytes read and written \\
L1/L2   & L1TEX or L2 access bytes \\
ALU   & Arithmetic/logic instructions \\
FMA   & Fused multiply-add instructions \\
FP64  & Double-precision (FP64) instructions \\
Mem   & Memory access instructions \\
Ten   & Tensor core instructions \\
\bottomrule
\end{tabular}
\label{tab:metrics}
\end{table}

\subsection{What-if Analysis on DAG}
\label{whatif}
Using the information collected by the trace profiler, we construct an execution graph in which each execution primitive is represented as a node, and dependencies are represented as directed edges. For the what-if analysis, our goal is to determine how frequency changes in GPU compute kernels or packs affect end-to-end latency. Formally, the execution graph is a directed acyclic graph (DAG), denoted as $G = (V, E)$, where each node $v \in V$ corresponds to an event, and each edge $(i, j) \in E$ encodes a precedence constraint: event $i$ must finish before event $j$ can begin. This DAG structure captures both the parallelism and dependency relationships within the workload. For each node $v$, we define its completion time as:
\begin{align}
\label{vtime}
    C_v = t_v + \max_{k\in \mathrm{pred}(v)} C_k,
\end{align}
where $t_v$ is the execution time of node $v$ measured during trace profiling, and $\mathrm{pred}(v)$ denotes the set of predecessor nodes for $v$.
The end-to-end latency and total energy consumption of the workflow are then given by:
\begin{align}
\label{e2etime}
    T_{\mathrm{DAG}} = \max_{v\in V} C_v, \quad E_{\mathrm{DAG}} = \sum_{v\in V} e_v + E_{\text{idle}},
\end{align}
where $e_v$ is the energy consumed by node $v$, and $E_{\text{idle}}$ represents idle energy consumption.

To optimize $E_{\mathrm{DAG}}$ and $T_{\mathrm{DAG}}$, we consider adjusting the GPU frequency. When the operating frequency changes, both $t_v$ and $e_v$ become functions of the frequency. For each node:
\begin{align}
    t_v \rightarrow t_v^{(f_v)}, \quad e_v \rightarrow e_v^{(f_v)},
\end{align}
where $f_v$ is the operating frequency during the execution of node $v$. Note that for non-GPU-kernel nodes, the execution time or energy may not be affected by GPU frequency.

We define a frequency policy $P = \{\,\langle T_i, F_i\rangle \mid $ $i = 1,2,...\,\}$, where $T_i$ are the time points at which frequency changes occur, and $F_i$ are the corresponding frequency settings. For simplicity, we omit GPU IDs, although in practice, each GPU may be governed by its own policy. The optimization objective is to find a feasible policy $P$ that minimizes total energy consumption, subject to an end-to-end latency constraint:

\begin{align}
\begin{aligned}
\label{optproblem}
\arg\min_{P}\quad\ &\texttt{Energy} = \sum_{v\in V} e_v^{(f_v)} + E_{\text{idle}} \\
\text{s.t.}\quad\ 
& T_{\mathrm{DAG}} = \max_{v\in V} C_v \leq T_{\mathrm{set}}, \\
& C_v = t_v^{(f_v)} +  \max_{k\in \mathrm{pred}(v)} C_k, \quad \forall v\in V, \\
& \Delta T_i = T_{i+1} - T_i \geq \epsilon, \quad 1\leq i\leq |P|, \\
% & |P| \leq M,
\end{aligned}
\end{align}
where
\begin{align*}
    f_v = F_i \quad \text{if} \quad \max_{k\in \mathrm{pred}(v)} C_k \in [T_i,T_{i+1}].
\end{align*}
This formulation provides a general objective to guide energy optimization under performance constraints. When the frequency policy $P$ changes, the frequency $f_v$ of node $v$ changes accordingly. This alters the node's duration $t_v^{(f_v)}$. If this change causes node $v$ to fall on the critical path, the end-to-end execution time $T_{\mathrm{DAG}}$ will shift.

\subsection{Packing Strategies}
\label{analysis}
Kernel packing essentially aggregates the GPU compute kernels into larger units in the DAG. While kernel packing effectively reduces profiling overhead, it also disrupts the original execution graph's dependencies. This disruption can potentially invalidate the end-to-end performance analysis described in Equation~\ref{e2etime}.

The DAG's end-to-end latency is determined by each node's critical predecessor $k_v^*$, defined as:
\begin{align}
k_v^* = \arg\max_{k \in \mathrm{pred}(v)} C_k,
\end{align}
where the edge $k_v^* \rightarrow v$ represents a critical dependency, and $v$ becomes the critical successor of $k_v^*$.

\begin{figure}[!htbp]
\centering
\includegraphics[width = 0.49\textwidth]{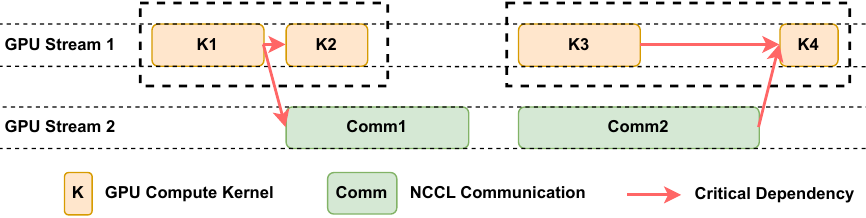}
\caption{Incorrect packing strategy leads to loss of critical dependencies.}
\label{packing2}
\end{figure}

When kernels are packed together, only the incoming dependencies of the first kernel and the outgoing dependencies of the last kernel are preserved. All intermediate dependencies are lost. As illustrated in Figure~\ref{packing2}, incorrect packing leads to the loss of critical dependencies. For instance, if we pack K3 and K4, we lose the critical dependency from K3 to Comm2. This loss results in incorrect critical path analysis and inaccurate performance predictions.

\lstset{
    backgroundcolor=\color{backcolour},   
    commentstyle=\itshape\color{codegreen},
    keywordstyle=\color{magenta},
    numberstyle=\color{codegray},
    stringstyle=\color{codepurple},
    basicstyle=\ttfamily\small,
    breakatwhitespace=false,         
    breaklines=true,                 
    captionpos=b,                    % Caption at the bottom
    keepspaces=true,                 
    numbers=left,                    
    numbersep=5pt,                  
    showspaces=false,                
    showstringspaces=false,
    showtabs=false,                  
    tabsize=2,
    xleftmargin=1.5em,                 % Add padding on the left
    xrightmargin=1em,                % Add padding on the right
    frame=none,                      % No hard frame, use background color instead
    escapeinside={(*@}{@*)},
    mathescape=true,
    framerule=0pt,
}
\begin{lstlisting}[language=Python, caption={Dependency-Aware Greedy Packing}, label={lst:packing}]
# Input: V (Sorted GPU kernels)
packs, curr_pack = [], []
for v in V:
    if not curr_pack:
        curr_pack = [v]; continue
        
    # Last node in current pack    
    u = curr_pack[-1] 
    
    # Dependency Safety 1
    r1 = (stream(v)==stream(v.crit_pred))
    # Dependency Safety 2
    r2 = all(stream(k)==stream(u) \
             for k in u.crit_succ)
    # Granularity Control 
    r3 = duration(curr_pack) + t_v < $\Delta_{\max}$
    
    # Check Rules
    if r1 and r2 and r3:
        curr_pack.append(v)
    else:
        packs.append(curr_pack)
        curr_pack = [v]

packs.append(curr_pack)
\end{lstlisting}

Our goal is to design a packing strategy ensuring that these lost dependencies would never become critical dependencies under any frequency setting. Based on the property that GPU compute kernel execution time monotonically increases as frequency decreases, we propose a \textbf{Dependency-Aware Greedy Packing} approach governed by the following three rules:

\begin{enumerate}[topsep=3pt, left=5pt, itemsep=1pt,parsep=0pt]
    \item \textbf{Dependency Safety 1:} For all non-initial nodes $i$ within a pack, ensure that $i$ and its critical predecessor $k_i^*$ belong to the same stream.
    \item \textbf{Dependency Safety 2:} For all non-final nodes $i$ within a pack, ensure that $i$ and all its critical successors belong to the same stream.
    \item \textbf{Granularity Control:} The cumulative execution time of a pack under the default frequency should not exceed a predefined threshold $\Delta_{\max}$.
\end{enumerate}

The first two rules preserve key dependencies (Figure~\ref{packing2}), while the third ensures appropriate DVFS control granularity. Since GPU frequency adjustments typically require 5--15 ms to take effect, setting $\Delta_{\max} = 5$ ms balances control resolution with packing benefits. It is worth noting that because the DVFS transition latency is explicitly modeled in the constraints of Equation~\ref{optproblem}, the pack duration does not need to strictly align with this latency. Listing~\ref{lst:packing} details the procedure where we iterate through kernels in topological order and merge them only if these constraints are satisfied.

To further reduce overhead, we employ pack-level sampling by hashing the sequence of kernel names within each pack. Unlike individual kernels, whose performance varies significantly based on context and cannot be easily clustered~\cite{10.1145/3725843.3757107}, packs encapsulate sufficient contextual information to allow reliable sampling. Consequently, we profile only the first occurrence of each unique pack. These strategies collectively guide the pack profiler in acquiring accurate performance metrics for the modeling described in the subsequent section.

\subsection{Pack Modeling}
\label{sec:pack modeling}
Unlike general performance prediction models that estimate execution metrics from scratch~\cite{kandiah_accelwattch_2021}, our modeling task is much more focused: we already possess the exact characteristics of a pack at the maximum (default) frequency. Our goal is to predict its scaling behavior at lower frequencies. 

However, machine learning-based approaches~\cite{alavani_program_2023} are infeasible for this task because acquiring ground-truth energy labels at the pack level is impractical. GPU power sensors have a low sampling frequency (e.g., \(\sim\)50ms)~\cite{guerreiro_modeling_2019}, making it impossible to directly measure millisecond-scale packs. Measuring hundreds of looped packs to accumulate measurable energy is also inaccurate, since isolating real workload packs from live execution destroys runtime context and memory states.

These limitations lead us to adopt a simplified yet highly accurate microbenchmark-based characterization methodology established in prior studies~\cite{guerreiro_modeling_2019,guerreiro_gpgpu_2018,siesta,10.1145/3786205}. 
We use a combination of microbenchmarks as a proxy for each original pack, shifting the measurement burden from packs to microbenchmarks, which can be easily isolated and repeated for precise ground-truth energy profiling. 
This approach incurs only a \textbf{one-time offline overhead} while enabling rapid online predictions using the metrics collected in Table~\ref{tab:metrics}.

Specifically, we utilize a suite of 12 microbenchmarks. Building upon Guerreiro et al.~\cite{guerreiro_modeling_2019}, we augment standard scalar and memory benchmarks with specialized Tensor Core workloads (HMMA and IMMA) to capture the matrix-heavy nature of modern AI models. We profile the power and latency of these microbenchmarks offline across all frequency states to construct a metric matrix $\mathbf{B} \in \mathbb{R}^{n \times m}$.

At runtime, we characterize the target pack by collecting a feature vector $\mathbf{b} \in \mathbb{R}^n$ at the default frequency. We then decompose the pack's behavior into a linear combination of our microbenchmarks by solving a non-negative least squares (NNLS) problem~\cite{siesta} to find the composition weights $\mathbf{x} = (x_1, \ldots, x_m)^\top \in \mathbb{R}^m$:
\begin{align}
    \arg\min_{\mathbf{x}} \|\mathbf{B} \mathbf{x} - \mathbf{b}\|^2 \quad \text{s.t.} \quad x_i \geq 0, \quad \forall i \in \{1, \ldots, m\}.
\end{align}

With the weights $\mathbf{x}$ resolved, the predicted energy $E_f$ and latency $T_f$ of the pack at any target frequency $f$ are efficiently computed as the weighted sum of the pre-profiled microbenchmark metrics:
\begin{align}
    \begin{aligned}
    E_f[\texttt{P}] & = \sum_{i=1}^{m} x_i \cdot E_f[\texttt{B}_i], \quad
    T_f[\texttt{P}] & = \sum_{i=1}^{m} x_i \cdot T_f[\texttt{B}_i].
    \end{aligned}
\end{align}
This lightweight analytic model allows EMO to instantly derive frequency scaling behaviors, decoupling energy modeling from online power measurements.

\begin{algorithm}
\caption{End-to-End Prediction (What-If Analysis)}
\label{alg:sim}
\begin{algorithmic}[1]
\REQUIRE DAG $G=(V,E)$; frequency policy $P$; \\
     Modeling tables $T_f, E_f$ for packs; idle power $p_\text{idle}[f]$
\ENSURE End-to-end time $T$, total energy $E$
\STATE Initialize $comp[v]$, $pred[v]$ for all $v \in V$
\STATE $E_a \gets 0$, $E_i \gets 0$ \quad \texttt{// active and idle energy}
\FOR{node $v$ in topological order of $G$}
    \STATE $start \gets comp[pred[v]]$
    \IF{$v$ is a pack}
        \STATE $f \gets F_i$ where $start \in [T_i,T_{i+1})$ in $P$
        \STATE $dur \gets T_f[v]$,\quad $e \gets E_f[v]$
    \ELSE
        \STATE $dur\gets t_v$,\quad $e\gets 0$
    \ENDIF
    \STATE $comp[v] \gets start + dur$,\quad $E_a \gets E_a + e$
    \FOR{each $u$ in successors of $v$}
        \IF{! $pred[u]$ \textbf{or} $comp[v] > comp[pred[u]]$}
            \STATE $pred[u] \gets v$
        \ENDIF
    \ENDFOR
\ENDFOR

\FOR{each $[T_i,T_{i+1})$ in $P$}
    \STATE $E_i \gets E_i + \texttt{idle\_time}_{[T_i,T_{i+1})} \times p_{\text{idle}}[F_i]$
\ENDFOR

\STATE $T \gets \max_{v \in V} comp[v]$
\STATE $E \gets E_a + E_i$
\RETURN $T, E$
\end{algorithmic}
\end{algorithm}

\subsection{Energy Efficiency Optimization}

We have formulated frequency policy optimization as the problem in Formulation~\ref{optproblem}. To efficiently evaluate optimization objectives, we integrate the pack-level models into the global execution graph to perform a holistic what-if analysis.

Given a frequency policy, we can efficiently predict end-to-end execution using our performance models and execution graph. Algorithm~\ref{alg:sim} details this prediction process: we perform a single topological traversal of the DAG, updating each pack's time and energy consumption under its assigned frequency while tracking critical predecessors. We then account for idle energy consumption to obtain end-to-end predictions. Since this requires only one topological traversal, the time complexity is $O(|V|+|E|)$.

\begin{algorithm}
\caption{Search Near-optimal Frequency Policy}
\label{alg:dp}
\begin{algorithmic}[1]
\REQUIRE
    frequency intervals $\{[p_i,p_{i+1})\}$; constraint $T_{\max}$; \\
    supported frequencies $\{F\}$; Predict function $\mathrm{Pred}(P)$; \\
\ENSURE
    $P^*$: near-optimal policy
\STATE Initialize $P_0$ = default freq in all intervals
\STATE ($t_0$,$E_0$) = $\mathrm{Pred}(P_0)$
\STATE Initialize $dp[0][t_0] \gets (E_0, P_0)$ 
\FOR{$i = 0$ to $|P|-1$}
    \FOR{each $(t, (E, P))$ in $dp[i]$}
        \FOR{each $f \in F$}
            \STATE $P' \gets$ set freq $f$ in $[p_i, p_{i+1})$ in $P$
            \STATE $(t', E') \gets $ $\mathrm{Pred}(P')$
            \IF{$t' \leq T_{\max}$ \AND $E' < dp[i+1][t'].\text{energy}$}
                \STATE $dp[i+1][t'] \gets (E', P')$
            \ENDIF
        \ENDFOR
    \ENDFOR
    \STATE Clear $dp[i-1]$ if $i > 0$\quad \texttt{// memory recycle}
\ENDFOR
\STATE $(E^*, P^*) \gets$ $\min_{t\leq T_{max}} dp[N][t]$
\RETURN $P^*$
\end{algorithmic}
\end{algorithm}

\begin{figure*}[!htbp]
\centering
\includegraphics[width = 0.90\textwidth]{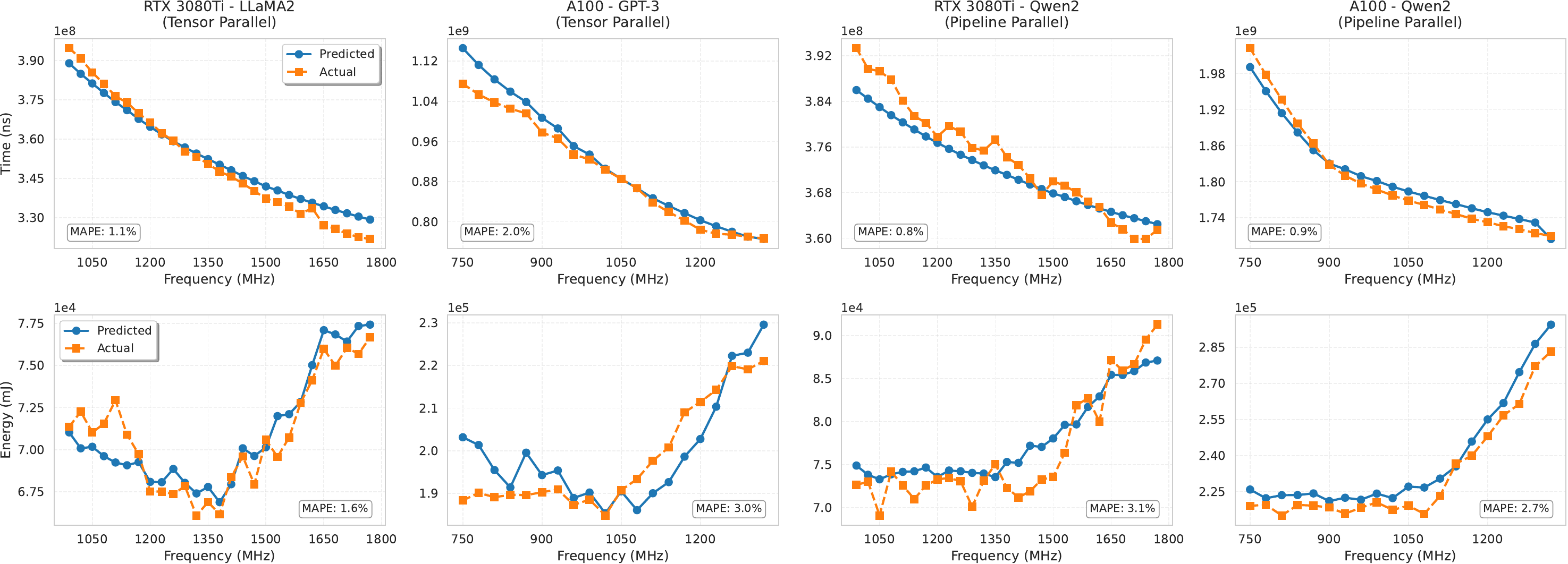}
\caption{Performance and energy modeling accuracy across different workloads and GPU configurations. The top row shows execution time predictions with MAPE, while the bottom row displays energy consumption predictions. EMO achieves low prediction errors across diverse scenarios.}
\label{modeling}
\end{figure*}

However, finding the optimal policy is challenging. Even coarse-grained versions of this problem are NP-hard~\cite{chung_reducing_2024}. Therefore, we introduce several simplifications to enable efficient search. The energy optimization problem on graphs shares similarities with the knapsack problem~\cite{chung_reducing_2024}. Inspired by this connection, we develop a pseudo-polynomial time algorithm. We first discretize the performance constraint and fix the frequency change time points $T_i$. This discretization introduces minimal error since we can set an arbitrarily fine granularity (e.g., 1 ms), and our problem formulation already imposes minimum interval constraints on frequency transitions.

As shown in Algorithm \ref{alg:dp}, we employ dynamic programming with a state array $dp[i][t]$ storing the minimum energy consumption when the first $i$ frequency change points have been processed and the end-to-end time is $t$. This is a pseudo-polynomial algorithm, as the time complexity is proportional to $\Delta T = T_{\text{set}} - T_0$, where $T_0$ is the end-to-end latency at maximum frequency (frequency scaling can only increase execution time). Since each state update requires invoking the prediction algorithm, the total time complexity is $O(|P| \cdot |F| \cdot \Delta T \cdot (|V|+|E|))$, where $|P|$ is the number of frequency change points and $|F|$ is the number of available frequency levels.

\section{Evaluation}
\subsection{Setup}

\textbf{Workloads and Hardware.} 
We evaluate EMO across workloads that demonstrate the three asynchronous patterns from Section \ref{sec:key-insights}. 
First, we evaluate tensor-parallel training via DeepSpeed Domino~\cite{wang2024-deepspeed-domino} for computation-communication overlap. 
Second, we examine pipeline-parallel training with DeepSpeed~\cite{10.1145/3394486.3406703} for inter-GPU asynchrony. 
Third, we test KTransformers~\cite{ktrans} inference with Mixture-of-Experts (MoE) offloading for CPU-GPU asynchrony.

Our hardware testbed includes two NVIDIA RTX 3080 Ti GPUs (consumer-grade), and eight A100-80GB GPUs (datacenter). On the 3080 Ti platform, due to memory constraints, we train LLaMA2-0.5B and Qwen2-0.5B~\cite{yang2024qwen2technicalreport}. 
On the A100 cluster, we train GPT3-6.7B~\cite{GPT3} and Qwen2-7B~\cite{yang2024qwen2technicalreport}, and perform inference on DeepSeek-V2-Lite~\cite{deepseekai2024deepseekv2strongeconomicalefficient}.

\textbf{Frequency Control and Energy Monitoring.}
We employ Zeus's daemon process (zeusd)~\cite{you_zeus_nodate} for frequency management, avoiding the need for administrative privileges.
While our method supports both memory and SM/core frequency scaling, A100 GPUs restrict memory frequency adjustment, so we focus on SM frequency control in our experiments.
Energy is monitored via NVIDIA Management Library, reporting 5-iteration averages to account for measurement variability.

\textbf{Baselines.}
We compare with both model-free and model-based approaches. Model-free baselines include Zeus~\cite{you_zeus_nodate}, which optimizes energy using power limits, and GEEPAFS~\cite{zhang_improving_2024}, which employs online modeling with GPU memory bandwidth utilization. For model-based comparison, we use Perseus~\cite{chung_reducing_2024} in pipeline-parallel scenarios, as it represents the state-of-the-art in phase-aware frequency scaling. Other relevant works, such as Wang et al.~\cite{wang_using_2025} and EVeREST~\cite{yue_everest_2025}, are omitted due to the lack of open-source implementations.

\begin{figure}[!htbp]
\centering
\includegraphics[width = 0.49\textwidth]{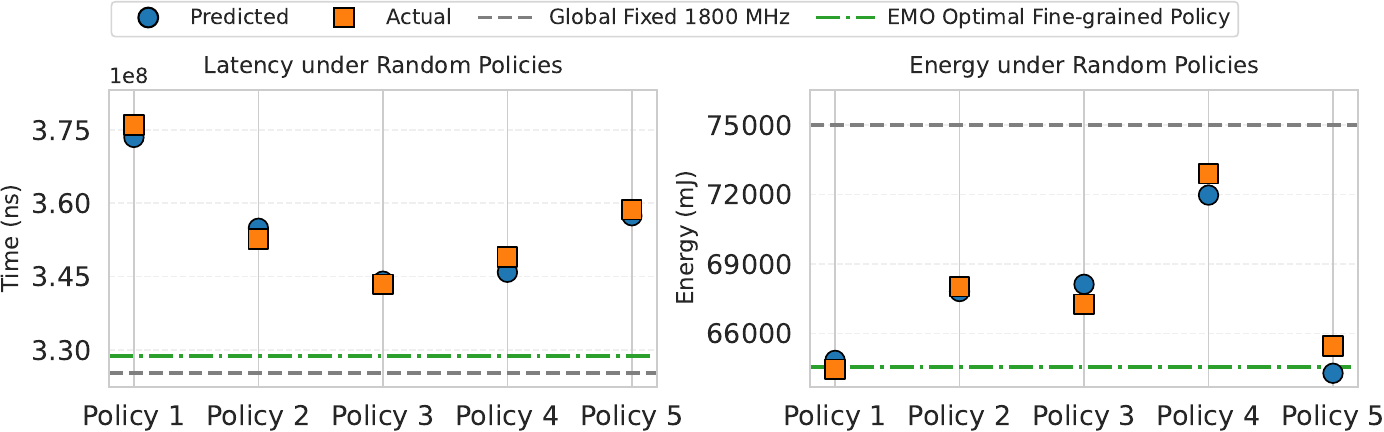}
\caption{Prediction accuracy of latency and energy under fine-grained random frequency policies for LLaMA2 with Tensor Parallelism on RTX 3080 Ti.}
\label{random}
\end{figure}

\subsection{Performance and Energy Modeling}
\label{eval:modeling}

As discussed in Section \ref{sec:pack modeling}, directly validating runtime pack-level energy is impractical due to the low sampling frequency of GPU power sensors. Consequently, we validate EMO's fidelity through end-to-end measurements. Unlike individual packs, end-to-end execution spans a sufficient duration for GPU power sensors to capture reliable ground-truth energy. 

Specifically, our evaluation aims to verify whether EMO can accurately predict the end-to-end execution time \(T\) and total energy \(E\) under any given frequency policy \(P\). By feeding a predefined policy \(P\) into Algorithm~\ref{alg:sim}, EMO aggregates all pack-level predictions to estimate the global metrics, which are then compared against actual hardware measurements.

We first evaluate static policies across diverse workloads (LLaMA2, GPT-3, and Qwen2) on RTX 3080Ti and A100 platforms, where the entire workload runs at a fixed frequency $F$ (i.e., $P = \{ \langle 0, F \rangle \}$). As shown in Figure~\ref{modeling}, EMO achieves a Mean Absolute Percentage Error (MAPE) below 3\% for most configurations. Notably, it accurately captures the non-linear trend where energy sometimes increases at lower frequencies due to prolonged execution time ($E = Power \cdot t$). 

To further validate pack-level precision, we consider  
a highly complex scenario with fine-grained varying frequencies.
As shown in Figure~\ref{random}, 
we evaluate five offline-generated random policies.
In each policy, the GPU frequency is reassigned every 20ms to a random level between 900 MHz and 1800 MHz. 
Despite this complexity, EMO's end-to-end predictions remain highly consistent with actual measurements. Since the end-to-end prediction strictly aggregates all pack-level models via Algorithm~\ref{alg:sim}, this high accuracy under rapid frequency transitions proves the reliability of the underlying pack models.

\subsection{End-to-End Energy Optimization}

We quantify effectiveness using Energy Savings ($ES$) and Performance Loss ($PL$). Let $E_{ori}$ and $T_{ori}$ be the energy and latency at default frequency, and $E_{opt}$ and $T_{opt}$ be the metrics under optimized policies. They are calculated as:
\begin{equation}
    ES = \frac{E_{ori} - E_{opt}}{E_{ori}} \times 100\%, \quad PL = \frac{T_{opt} - T_{ori}}{T_{ori}} \times 100\%.
\end{equation}

\begin{table*}[htbp]
\centering
\caption{Comparison of various energy optimization methods for tensor-parallel training workloads}
\renewcommand{\arraystretch}{1.2}
\begin{tabular}{
    >{\centering\arraybackslash}p{2.0cm} 
    >{\centering\arraybackslash}p{2.7cm} 
    >{\raggedleft\arraybackslash}p{1.6cm}
    >{\raggedleft\arraybackslash}p{2.3cm}
    >{\raggedleft\arraybackslash}p{2.3cm}
    >{\raggedleft\arraybackslash}p{1.5cm}
    >{\raggedleft\arraybackslash}p{1.5cm}
}
\toprule
\textbf{Model Config} & \textbf{Metrics} & \textbf{ZEUS-5\%} & \textbf{GEEPAFS-2\%} & \textbf{GEEPAFS-5\%} & \textbf{EMO-2\%} & \textbf{EMO-5\%} \\
\midrule
\multirow{2}{*}{\shortstack{TP-2-LLaMA2 \\ On 3080 Ti}}
    & Energy Saving    & 14.72\% & 16.97\% & 19.41\% & 19.94\% & 27.56\% \\
    & Performance Loss & 4.82\% & 1.87\%  & 6.20\%& 1.33\% & 4.84\% \\
\midrule
\multirow{2}{*}{\shortstack{TP-2-GPT3 \\ On A100}}
    & Energy Saving   & 2.19\% & 0.40\% & 1.45\% & 7.00\% & 8.19\% \\
    & Performance Loss & 4.56\% & 2.59\%& 4.87\% & 2.36\% & 4.15\% \\
\midrule

\multirow{2}{*}{\shortstack{TP-4-GPT3 \\On  A100}}
    & Energy Saving   & 5.15\% & 0.69\% & 3.76\% & 12.08\% & 15.98\% \\
    & Performance Loss & 5.64\% & 2.80\%& 5.63\% & 1.99\% & 5.37\% \\
\midrule

\multirow{2}{*}{\shortstack{TP-8-GPT3 \\ On A100}} 
    & Energy Saving    & 2.31\% & 9.65\% & 14.16\% & 15.82\% & 20.53\% \\
    & Performance Loss & 3.21\% & 2.77\% & 4.21\% & 3.20\% & 4.78\% \\
\bottomrule
\end{tabular}
\label{tab:tp}
\end{table*}

\begin{table*}[htbp]
\centering
\caption{Energy savings (\%) comparison for pipeline-parallel training. Numbers in parentheses indicate actual performance loss (\%).}
\renewcommand{\arraystretch}{1.2}
\begin{tabular}{
    >{\centering\arraybackslash}m{2.5cm}
    >{\centering\arraybackslash}m{3.2cm}
    >{\raggedleft\arraybackslash}m{2.0cm}
    >{\raggedleft\arraybackslash}m{2.5cm}
    >{\raggedleft\arraybackslash}m{2.5cm}
    >{\raggedleft\arraybackslash}m{2.3cm} 
}
\toprule
\textbf{Model Config} & \textbf{Partition Strategy} & \textbf{ZEUS} & \textbf{GEEPAFS} & \textbf{PERSEUS}& \textbf{EMO} \\
\midrule

\multirow{2}{*}{PP-2-QWEN2}
&Uniform&0.40 (4.91)  &4.22 (3.78) &6.06 (3.92)& 8.86 (0.28)\\
&Parameter&0.22 (2.78) &5.60 (1.71) &6.54 (3.25)& 14.42 (1.05)\\
% \cline{1-5}
\midrule
\multirow{2}{*}{PP-4-QWEN2}
&Uniform& 0.33 (2.04) &5.80 (3.73) &11.38 (4.39)& 18.97 (1.78)\\
&Parameter& 1.17 (2.04) &3.83 (4.69) &6.55 (4.29)& 16.22 (1.70)\\
% \cline{1-5}
\midrule
\multirow{2}{*}{PP-8-QWEN2}
&Uniform& 0.64 (1.45) &10.66 (5.38) &15.79 (7.84)& 18.89 (0.25)\\
&Parameter& 1.14 (2.56) &5.69 (3.14) & N/A & 17.17 (3.27)\\
\bottomrule
\end{tabular}
\label{tab:pp}
\end{table*}

\subsubsection{Tensor Parallel and Intra-GPU Asynchrony}
Tensor Parallelism (TP) splits operators across GPUs, requiring frequent communication. The main optimization opportunity here is intra-GPU asynchrony: the overlap between computation and communication on different CUDA streams. The degree of this overlap determines the execution slack.
Table~\ref{tab:tp} presents the energy optimization results for TP training workloads. EMO consistently outperforms both model-free approaches (GEEPAFS and Zeus) across different configurations. 

In scenarios with significant communication bottlenecks, such as TP-2 on the RTX 3080 Ti and TP-8 on the A100, the execution slack is substantial. EMO effectively identifies these non-critical periods to achieve 27.56\% and 20.53\% energy savings, significantly surpassing the savings achieved by GEEPAFS. Zeus underperforms here because it relies on power capping rather than the direct frequency scaling~\cite{yue_everest_2025}.

Conversely, configurations like TP-2-A100 exhibit minimal slack. Existing methods like GEEPAFS fail here because they rely on utilization drops for coarse-grained frequency scaling. Since GPU utilization remains high during communication waits, it cannot distinguish critical computation from non-critical communication, leading to excessive performance loss. 
EMO overcomes this limitation by exploiting intra-GPU asynchrony with millisecond granularity, making it highly effective for TP workloads that demand multi-phase, fine-grained optimization. In our experiments, we adjust frequencies every 20ms, a precision enabled by our high modeling accuracy.

\subsubsection{Pipeline Parallel and Inter-GPU Asynchrony}
We evaluate pipeline-parallel (PP) training on A100 GPUs using Qwen2-7B with 2, 4, and 8-GPU configurations. We utilize DeepSpeed's two partitioning strategies: uniform (equal layers per stage) and parameter-based (equal parameter size per stage). PP workloads exhibit inter-GPU asynchrony, where GPUs wait for neighbor data dependencies. These pipeline bubbles form the execution slack.
Table~\ref{tab:pp} compares EMO against baselines for PP training under strict performance constraints (targeting zero performance loss). EMO achieves up to 18.97\% energy savings for PP-4 (uniform) with minimal performance impact (1.78\%). Perseus, despite targeting PP, achieves only 11.38\% savings with higher variability (4.39\%).

This stems from EMO's superior bubble utilization. While both methods identify the execution slack created by bubbles, Perseus operates at a coarse-grained \textit{Pass Level}, applying a single frequency to an entire forward or backward stage. This is inefficient because operations within a pass have varying frequency sensitivities. Within the same fixed time budget (the bubble size), Perseus is limited by the most sensitive parts of the pass. In contrast, EMO operates at the \textit{Pack Level}, aggressively slowing frequency-insensitive packs while keeping sensitive ones fast, extracting greater savings.  

Furthermore, Perseus struggles with short pipeline stages, such as the extremely brief first stage in the 8-GPU parameter-based partitioning. Relying on online measurements, it encounters the hardware limitation discussed in Section~\ref{eval:modeling}: GPU power sensors sample too slowly to profile such short phases, causing Perseus to fail (marked as N/A). In contrast, EMO completely bypasses online energy measurement and generates robust policies even for highly imbalanced pipelines.

\subsubsection{Offloading and CPU-GPU Asynchrony}
\begin{figure}[!htbp]
\centering
\includegraphics[width = 0.48\textwidth]{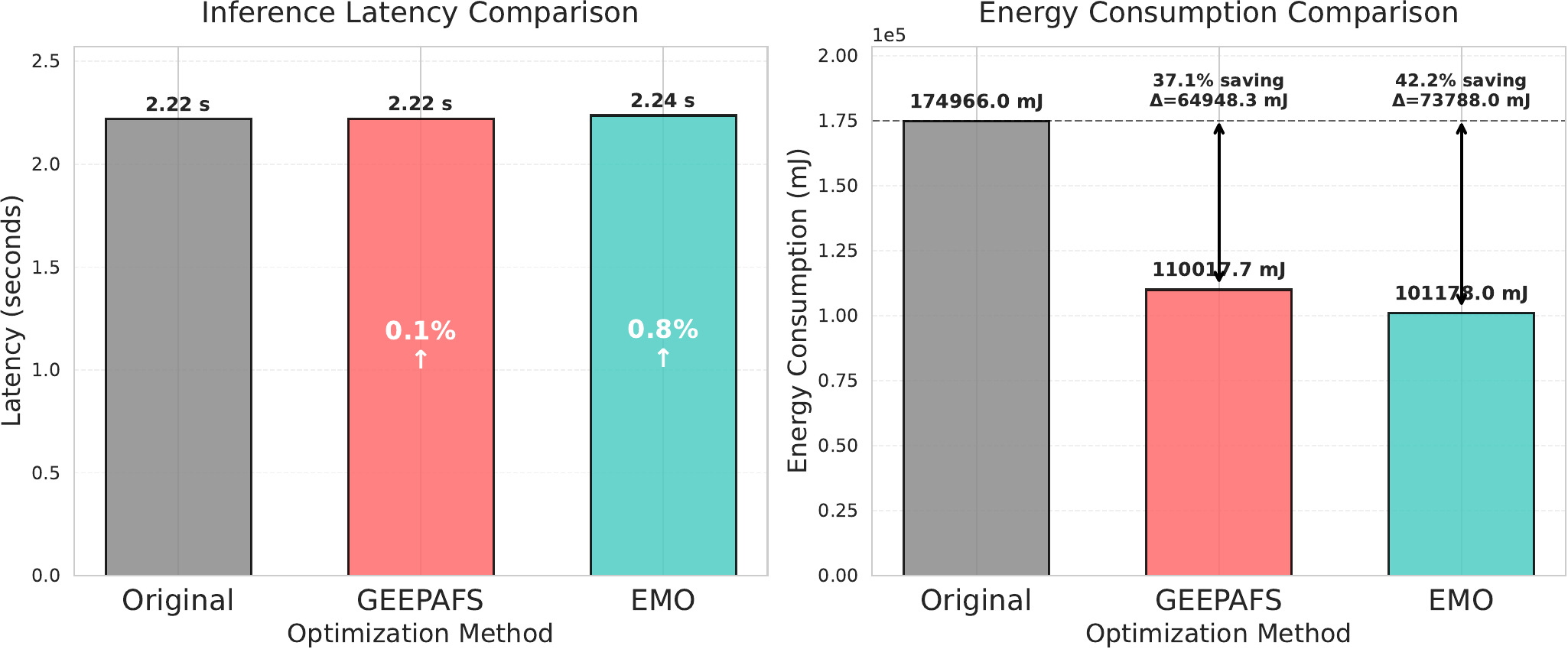}
\caption{Energy optimization results for DeepSeek-V2-Lite inference with KTransformers offloading. }
\label{inference}
\end{figure}

Figure~\ref{inference} shows optimization results for DeepSeek-V2-Lite inference via KTransformers. It enables inference under memory constraints by offloading MoE layers to the CPU, while GPUs handle attention computation. This setup creates a CPU-bottlenecked scenario with low GPU utilization. Consequently, reducing GPU frequency rarely impacts end-to-end performance. 

In this scenario, GEEPAFS's utilization-based approach also performs well, achieving results close to EMO. EMO achieves 42.2\% energy reduction with only 0.8\% latency increase, compared to GEEPAFS's 37.1\% energy savings with 0.1\% latency increase. EMO's slight advantage lies in its stable policies, whereas GEEPAFS requires periodic online exploration, a fundamental limitation of model-free methods. 

\subsection{Overhead Analysis}
\begin{figure}[!htbp]
\centering
\includegraphics[width = 0.48\textwidth]{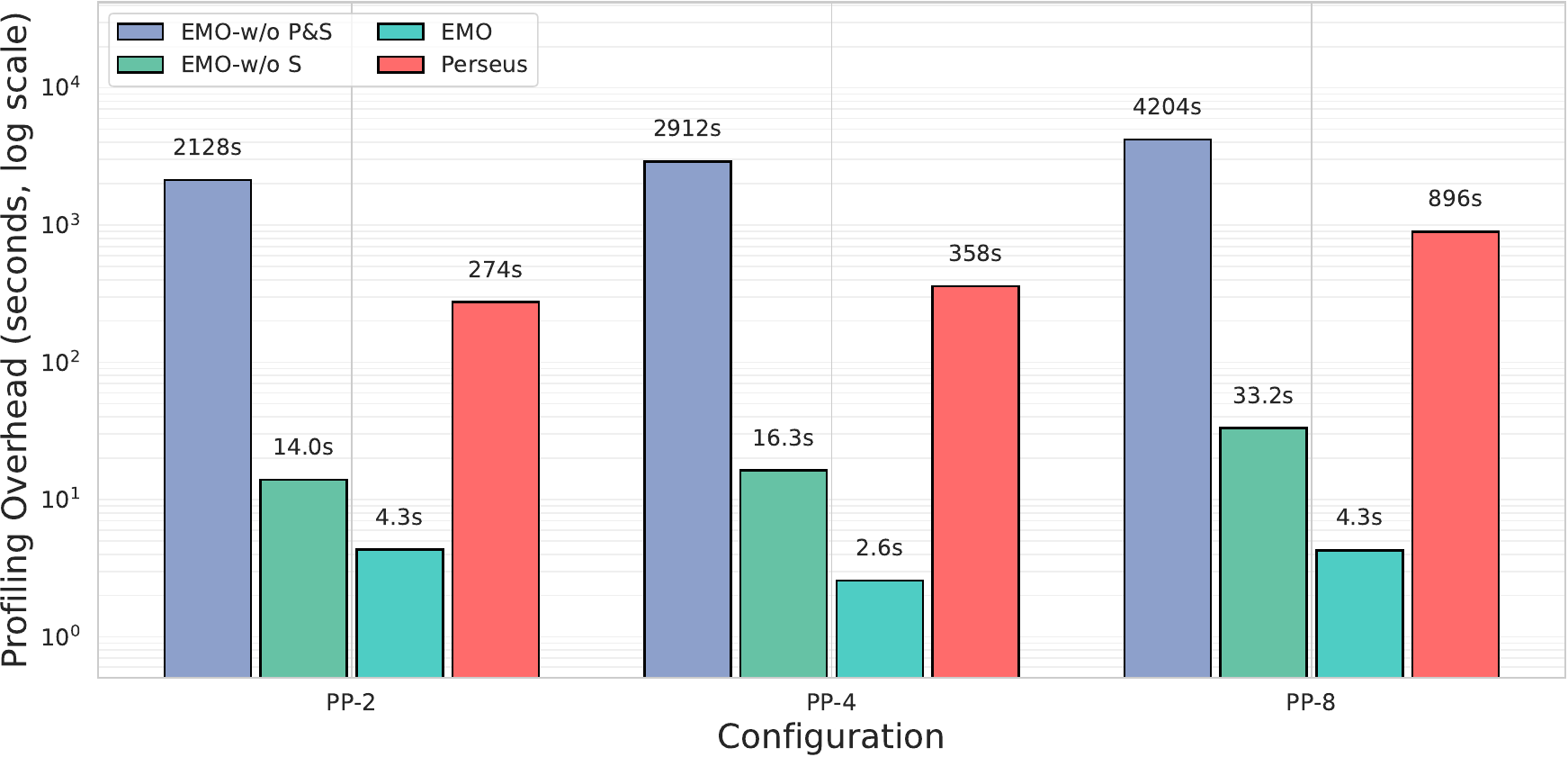}
\caption{Profiling overhead comparison on pipeline parallel workloads.}
\label{overhead}
\end{figure}

Figure~\ref{overhead} compares profiling overhead on A100 GPUs using PP workloads to benchmark against Perseus. Model-free baselines (ZEUS, GEEPAFS) are excluded here due to their negligible profiling costs.

As shown in the figure, 
Perseus incurs substantial overhead to generate stable lookup tables, despite modeling at coarse pass levels. In contrast, EMO maintains under 5s overhead across all configurations. Remarkably, despite its finer pack-level granularity, EMO profiles orders of magnitude faster.

Figure~\ref{overhead} also presents an ablation study to quantify the efficiency gains. Raw kernel profiling (EMO-w/o Packing and Sampling) is prohibitively expensive ($\ge$2000s). Packing (EMO-w/o Sampling) drastically reduces modeled instances, and sampling (EMO) further avoids redundant profiling. Consequently, EMO is up to $1100\times$ faster than raw profiling and $50\times$–$200\times$ faster than Perseus.

While model-free approaches offer lower overhead, they suffer from limited optimization and unstable convergence. EMO successfully balances accuracy with deployment practicality, making it suitable for production environments requiring minimal overhead.

\section{Discussion}
\subsection{Generalization}
While our evaluation focuses on three fundamental parallelism paradigms, EMO’s design is inherently extensible to more complex hybrid parallelization strategies and heterogeneous computing environments. This generality stems from our decision to model at the level of execution primitives rather than high-level framework operations. By abstracting the workload into a low-level dependency graph, EMO decouples itself from specific model architectures. Furthermore, constructing the graph via CUPTI tracing ensures compatibility with runtime optimizations like CUDA Graphs.

Although EMO currently targets stable patterns, execution variability is often structured and predictable. In LLM serving, graph topology remains highly similar across requests. Integrating output length prediction~\cite{kakolyris_throttllem_2025} could allow EMO to dynamically adapt policies without full re-profiling.

\subsection{Optimization Scalability}
Our pseudo-polynomial algorithm provides optimal solutions, but search time grows with GPU scale. Fortunately, exhaustive search is often unnecessary. In tensor- and data-parallel setups, we can optimize a single GPU's graph and replicate the policy. Only heterogeneous or pipeline-parallel workloads require multi-device graph merging.

Although optimization search in Algorithm \ref{alg:dp} runs asynchronously without consuming GPU resources, EMO can alternatively integrate approximation methods. For example, reinforcement learning~\cite{wang_drlcap_2024} or evolutionary search methods~\cite{wang_using_2025,10611913} can efficiently explore the solution space, trading strict optimality for bounded search times in large-scale deployments.

\section{Conclusion}
This paper presented EMO, a lightweight framework for energy optimization across diverse AI workloads. By introducing the execution graph abstraction, EMO captures fine-grained slack windows while maintaining workload generality. Our kernel packing techniques balance modeling accuracy and profiling overhead. Through pack modeling and intelligent graph analysis, EMO achieves up to 28\% energy savings with minimal performance impact across diverse AI workloads.

\section*{Acknowledgment}
We utilized a generative AI tool to assist in improving the
grammar and readability of this manuscript.

\bibliographystyle{IEEEtran}
\bibliography{sample}

\end{document}